\documentclass[aps,prd,twocolumn]{revtex4}

\usepackage{graphicx}

\newcommand \tg{\tilde\gamma}
\newcommand \tG{\tilde\Gamma}
\newcommand \tA{\tilde A}
\newcommand \p{\partial}
\newcommand \pn{\partial_n}
\newcommand \g{\gamma}

\newcommand \re{{\rm Re}\,}

\begin{document}

\title{Symmetric hyperbolicity and consistent boundary conditions for
second-order Einstein equations}

\author{Carsten Gundlach}
\email[]{C.Gundlach@maths.soton.ac.uk} 
\affiliation{School of Mathematics, University of Southampton,
         Southampton SO17 1BJ, UK}

\author{Jos\'e M. Mart\'\i n-Garc\'\i a}
\email[]{jmm@imaff.cfmac.csic.es} 
\affiliation{\mbox{Instituto de Matem\'aticas y F\'{\i}sica Fundamental,
Centro de F\'{\i}sica Miguel A. Catal\'an,} \\
C.S.I.C., Serrano 113 bis, 28006 Madrid, Spain}

\date{3 March 2004}


\begin{abstract}

We present two families of first-order in time and second-order in
space formulations of the Einstein equations (variants of the
Arnowitt-Deser-Misner formulation) that admit a complete set of
characteristic variables and a conserved energy that can be expressed
in terms of the characteristic variables. The associated constraint
system is also symmetric hyperbolic in this sense, and all
characteristic speeds are physical. We propose a family of
constraint-preserving boundary conditions that is applicable if the
boundary is smooth with tangential shift. We conjecture that the
resulting initial-boundary value problem is well-posed.

\end{abstract}


\maketitle


\section{Introduction}


Current numerical relativity codes designed to simulate the inspiral
and merger of a black hole binary are limited by instabilities. These
are now believed to be usually instabilities of the continuum
equations, rather than of the numerical method. Intuitively one feels
that the initial value problem for the Einstein equations is
well-posed in a geometrical sense. For instance, in fully harmonic
spacetime coordinates the Einstein equations reduce to ten quasilinear
wave equations \cite{Wald}. However, in the usual 3+1 split one uses
only six of the ten Einstein equations for evolution, while the other
four must be imposed as constraints on the initial data. An evolution
in which the constraints are not obeyed is not a solution of the
Einstein equations.

Consider perturbing a solution of the evolution equations around a
solution of the full Einstein equations. We call a linear perturbation
physical if it obeys the constraints initially and therefore at all
times, and if it cannot be removed by a change of coordinates. A
perturbation is called pure gauge if it obeys the constraint but can
be removed by a coordinate transformation. A perturbation is called
unphysical if it violates the constraints. All of these perturbations
can be instabilities in the (weak) sense that they
grow with respect to the background solution. An example for a
physical instability is one that pushes a marginally stable star over
the edge of gravitational collapse. An example of a gauge instability
is given by the evolution of Minkowski spacetime with constant lapse,
with an initial slice that is not quite flat: the time slices become
singular in a finite time. Some, perhaps all, gauge instabilities can
be avoided by a suitable choice of the lapse and shift adapted to the
solution.

The problem in numerical solutions are unphysical instabilities. These
will always be triggered in a numerical simulation by finite
differencing or round-off error. Furthermore, the solution space of
the evolution equations is infinitely bigger than the solution space
of the full Einstein equations, and so the latter are likely to
represent an unstable equilibrium. Evolution schemes which replace
between one and four of the evolution equations with constraint
equations (constrained evolution) do not fundamentally address this
problem, as they are still only solving six equations.

In discussing instabilities, it is important to distinguish between
evolution systems which are well-posed and evolution systems which are
ill-posed. In the former, the growth of any linear perturbation
$\delta u(x,t)$ of a background solution $u_0(x,t)$ can be bounded as
\begin{equation}
||\delta u(\cdot,t)||\le f(t) \ ||\delta u(\cdot,0)||,
\end{equation}
where $f(t)$ depends on $u_0$ but is independent of $\delta
u(x,0)$. This means that the solution $u_0+\delta u$ depends
continuously on its initial data. By contrast, in an ill-posed system
no such bound $f(t)$ exists. Rather, the growth rate increases
unboundedly with the highest spatial frequency present in $\delta
u(x,0)$. These are instabilities in a rather stronger sense. They can
be gauge, but typically are constraint-violating.

A numerical simulation inherits all the instabilities already present
in the continuum. A good numerical scheme does not add any others, but
can never fix continuum instabilities. A crucial point is that in a
numerical evolution, the highest spatial frequency present is
effectively always the grid frequency, because it is generated by
finite differencing error even if the initial data are smooth.

As one increases the resolution in numerical solutions of an ill-posed
system, constraint violating instabilities start with smaller
amplitude (because they are initialized by finite differencing error),
but grow more rapidly (because the finer grid can represent a higher
spatial frequency). Therefore they don't converge away. At high enough
resolution and late enough time this will become apparent as a
breakdown of convergence for the entire solution, but one can detect
their presence before that by looking at a Fourier transform in space
\cite{CalabreseWH}. As these non-convergent instabilities are features
of the continuum system they cannot be suppressed by any consistent
numerical dissipation.

By contrast, the numerical solution of a well-posed scheme will in
general still be swamped by constraint-violating instabilities at late
time, but now these start with smaller amplitude and grow at the same
rate as one increases resolution. They therefore converge away. 

In practice, depending on the system, physical solution, runtime and
resolution, either convergent or non-convergent instabilities can be
the dominant source of error. In 3-dimensional simulations in
particular, the available resolution is quite limited, and the lack of
convergence may not become apparent. The Arnowitt-Deser-Misner (ADM)
formulation of the Einstein equations has for many years been the main
formulation used in 3-dimensional simulations, even though it is
ill-posed (weakly hyperbolic) \cite{NOR}. More recently a systematic
comparison of high-resolution, long-time evolutions for well-posed and
ill-posed (weakly hyperbolic) systems \cite{CalabreseWH} has
demonstrated that the breakdown of convergence is really inevitable,
and has revived interest in well-posed formulations of the Einstein
equations.

This interest has focused on first-order reductions of the Einstein
equations, because for general first order evolution systems useful
criteria for well-posedness are known \cite{GustafssonKreissOliger}. A
sufficient and necessary criterion for the initial value problem (with
no boundaries, or periodic boundaries) to be well-posed is strong
hyperbolicity: this roughly means that the system has a complete set
of characteristic variables with real speeds. Symmetric hyperbolicity 
is another criterion which implies strong hyperbolicity and can be
used to obtain a well-posed initial-boundary value problem: roughly
speaking it means that the principal part of the system admits a
conserved energy. Therefore a number of strongly or symmetric
hyperbolic first-order reductions of the ADM equations have been
suggested over the years to assure well-posedness, see for example
\cite{FrittelliReula,AndersonYork,KST,ReulaLivRev}.In a symmetric hyperbolic
formulation of the Einstein equations, the constraints can be dealt
with consistently in the initial-boundary value problem (see
\cite{FriedrichNagy} for the full, and
\cite{CalabreseCPBC,SzilagyiWinicour} for the linearised Einstein
equations).

However, first order reductions introduce new, auxiliary, variables
and constraints, and so further increase the solution space. One
would expect this to give rise to additional constraint-violating
instabilities of the convergent type, even if well-posedness rules out
the non-convergent type. There is some evidence that this is a real
problem \cite{KSTmodes}, which may even outweigh the benefits of
hyperbolicity. It would therefore seem preferable to find a system
that is symmetric hyperbolic while enlarging the solution space as
little as possible, and in particular this should be a second-order
system.

In a companion paper \cite{bssn1} we have proposed a definition of
symmetric hyperbolicity for second-order systems. This can be used to
obtain a well-posed initial-boundary value problem in the same way as
in first-order systems, but without enlarging the solution space. Here
we show symmetric hyperbolicity for two formulations of the Einstein
equations that are variants of the Arnowitt-Deser-Misner (ADM)
equations, namely the Baumgarte-Shapiro-Shibata-Nakamura (BSSN) system
\cite{SN,BS}, which is already widely used in numerical relativity
\cite{AlcubierreBSSN,AEIBBH,Shibata,LagunaBBH}, and a simpler related
system suggested by Nagy, Ortiz and Reula (NOR) \cite{NOR}. 

We know of two previous partial results concerning the hyperbolicity
of a second-order, ADM-like version of the Einstein equations. In
\cite{SarbachBSSN} it was shown that a first-order reduction of the
BSSN system is strongly hyperbolic, and a variant with some
superluminal characteristic speeds is symmetric hyperbolic. It was
then noted that the auxiliary constraints associated with the
introduction of the first-order auxiliary variables form a closed
subsystem of the constraint system. This means that if the auxiliary
constraints are obeyed initially, and if suitable boundary conditions
are imposed, they are obeyed during the evolution, even if the other
constraints are not obeyed. In this sense, the introduction of the
auxiliary variables has not enlarged the solution system. In
\cite{NOR} it was shown that the NOR system (in second-order form),
and its associated constraint system are strongly hyperbolic, by using
a pseudo-differential reduction to first order which also does not
enlarge the solution space. This method relies in an essential way on
Fourier transforms, and cannot be used to deal with the
initial-boundary value problem.

Here we go beyond these two papers in showing symmetric hyperbolicity
for the second order BSSN and NOR systems and their associated
constraint systems, without any reduction to first order. We do this
by defining characteristic variables, finding a conserved positive
definite covariant energy, and expressing it in terms of the
characteristic variables. We have chosen BSSN and NOR because BSSN is
popular with numerical relativists today, whereas NOR seems a simpler
version of BSSN that shares all its advantages.

The paper is organised as follows. In Sec.\ref{section:method} we
introduce our method and general notation. In Sec.~\ref{section:bssn}
we prove symmetric hyperbolicity for the BSSN evolution equations and
the associated constraint system, and in Sec.~\ref{section:nor} we do
the same for the NOR system. We take a first look at the initial
boundary problem for both systems in Sec.~\ref{section:boundary}. We
propose a family of boundary conditions for a smooth boundary where
the shift is tangential to the boundary. We show that there are no
arbitrarily rapidly growing modes --- this amounts to a strong
indication of well-posedness but is short of a
proof. Sec.~\ref{section:conclusions} contains our conclusions.


\section{Method and notation}
\label{section:method}


Here we briefly summarise the relevant notation and methods of
\cite{bssn1}. We need to split 3-tensors into their longitudinal
and transversal parts with respect to a given direction $n_i$. Assume
a 3-metric, say $\g_{ij}$, which will be used to raise and lower
indices, and let $n^i$ be a unit vector with respect to this
metric. Then
\begin{equation}
{q_i}^j\equiv {\delta_i}^j -n_i n^j
\end{equation}
is the projector into the space transversal to $n_i$. A tensor index
that has been projected will be denoted by the index $A,B,\dots$
instead of $i,j,\dots$. An index $n$ denotes a tensor index
contracted with $n_i$ or $n^i$.  A self-explanatory example of this
notation is
\begin{equation}
P^i\partial_i \equiv P_n \partial_n + P^A \partial_A.
\end{equation}
A pair $qq$ of indices indicates a contraction with $q_{ij}$, and for
tensors of rank 2 or higher, we use the convention that they are
totally tracefree on their projected indices.

Now consider a system of evolution equations that are second order in
space and first order in time. Linearise around a solution, and
approximate the background-dependent coefficients of the linearised
system as constant (frozen). Retain only the principal part of the
equations. We call the resulting linear system with constant
coefficients strongly hyperbolic if for any given direction $n^i$ we
can find a complete set of characteristic variables $U$ that obey
\begin{equation}
\partial_t U=\lambda \partial_n U + \hbox{transversal derivatives}.
\end{equation}
Here $-\lambda$ is the propagation speed in the $n_i$ direction.
The $U$ are constructed from $\tilde u\equiv (u,\partial_i u)$. This
definition has two important consequences: transversal derivatives
$\partial_A u$ are automatically zero speed variables, and
arbitrary multiples of transversal derivatives $\partial _A u$ can be
added to any characteristic variable. Below we shall find the
characteristic variables $U$ in two steps: we first find a set of
non-zero speed variables $U'$ that do not contain any transversal
derivatives, then add transversal derivatives to them until we can
express the energy and flux in terms of the modified characteristic
variables $U$.

We call the system symmetric hyperbolic if it admits an energy
\begin{equation}
E=\int_\Omega \epsilon \,dV,
\end{equation}
where $\epsilon$ is covariant, positive definite, and conserved in the
sense that
\begin{equation}
\partial_t \epsilon=\partial_i F^i
\end{equation}
for some flux $F^i$, {\it and} if we can express $\epsilon$ and $F^n$ in
terms of characteristic variables. 

The simplest example is the wave equation in the form
\begin{eqnarray}
\label{wave1}
\partial_t\phi&=&\Pi, \\
\label{wave2}
\partial_t\Pi&=&\partial_i\partial^i\phi,
\end{eqnarray}
with $u=(\phi,\Pi)$ and $\tilde u=(\partial_i\phi,\Pi)$.
The characteristic variables in a given direction $n_i$ are
\begin{eqnarray}
\label{2ndchar}
U_\pm &\equiv& \Pi\pm \partial_n \phi, \\
U_A &\equiv& \partial_A \phi, 
\end{eqnarray}
with speeds $\lambda=(\pm 1,0)$. The covariant
energy and flux are
\begin{eqnarray}
\epsilon&=&\Pi^2+\partial_i\phi\partial^i\phi, \\
F^i&=&2\Pi\partial^i \phi.
\end{eqnarray}
In terms of characteristic variables they are
\begin{eqnarray}
\epsilon&=&{1\over 2}(U_+^2+U_-^2)+U_A U^A, \\
F^n&=&{1\over 2}\left(U_+^2-U_-^2\right).
\end{eqnarray}
Therefore
\begin{equation}
\label{Edot}
{dE\over dt}=\int_{\partial\Omega} {1\over 2}\left(U_+^2-U_-^2\right)\, dS
\end{equation}
where $U_\pm$ are now the characteristic variables normal to the
boundary $\partial\Omega$.

The maximally dissipative boundary condition
\begin{equation}
U_+=\kappa U_-+f
\end{equation}
with $|\kappa|\le 1$ and $f$ a given function then guarantees that the
growth of $E$ is bounded by $f$ and that $E$ does not grow for
$f=0$.

To make the energy positive definite in $\phi$ itself rather than just
$\partial_i\phi$, one can add a term $\alpha^2\phi^2$ to $\epsilon$,
with $\alpha>0$ constant. With a maximally dissipative boundary
condition and $f=0$, $E$ is then bounded by $E(t)\le e^{\alpha t}E(0)$.

Finally, allow the linearised system to have variable coefficients
(from the non-constant background solution) and a non-principal
part. For a finite time interval, the energy is then still bounded as
$E(t)\le Ke^{\alpha t}$ for some constants $K$ and $\alpha$, and the
linearised initial-boundary value problem remains
well-posed. Therefore it is sufficient to establish well-posedness to
examine the principal part in the frozen coefficient approximation. On
an intuitive level, this is so because the purpose of well-posedness
is to rule out instabilities of the non-convergent type, which have a
growth rate that grows with spatial frequency. Well-posedness of the
linearised problem is a necessary condition for well-posedness of the
full non-linear (quasilinear) problem.


\section{The BSSN formulation of the Einstein equations}
\label{section:bssn}



\subsection{Field equations}


The BSSN formulation of the Einstein equations (without matter) is
obtained from the ADM form of the Einstein equations \cite{York},
\begin{eqnarray}
\partial_t \gamma_{ij} &=& {\cal L}_\beta \gamma_{ij} -2\alpha K_{ij},
\\ 
\nonumber \partial_t K_{ij} &=& {\cal L}_\beta K_{ij} -D_iD_j\alpha \\
&&+\alpha\left(R_{ij}-2K_{il}{K^l}_j+KK_{ij}\right), \\
H&\equiv &R-K_{ij}K^{ij}+K^2=0, \\
M_i&\equiv &D_j{K^j}_i-D_iK= 0,
\end{eqnarray}
by introducing the new variables 
\begin{eqnarray}
\tg_{ij}&\equiv&(\det\gamma)^{-1/3}\gamma_{ij}, \\
\tG^i&\equiv&\tg^{ij}\tg^{kl}\tg_{jk,l}, \\
\phi&\equiv&{1\over12}\ln\det\gamma, \\
\tA_{ij}&\equiv&(\det\gamma)^{-1/3}\left(K_{ij}-{1\over
3}\gamma_{ij}K\right).
\end{eqnarray}
In the remainder of this Section, indices are moved with $\tg_{ij}$
and its inverse $\tg^{ij}$. Generalising the BSSN equations, we
densitise the lapse $\alpha$ with the determinant of the metric as
\begin{equation}
\alpha=e^{6\sigma\phi}Q
\end{equation}
where $\sigma$ is a constant and now $Q$, rather than $\alpha$, is a
given function of the coordinates. The definition of the $\tG^i$ gives
rise to the differential constraint
\begin{equation}
G_i\equiv \tg_{ij}\tG^j-\tg^{jk}\tg_{ij,k} = 0.
\end{equation}
The definition of $\tA_{ij}$ gives rise to the
algebraic constraint
\begin{equation}
{T}\equiv \tg^{ij}\tA_{ij}=0,
\end{equation}
and from the definition of $\tg_{ij}$ we have the algebraic constraint
\begin{equation}
D\equiv \ln \det \tg=0.
\end{equation}

The BSSN equations are first order in time, second order in space, and
quasilinear. The principal part of the evolution equations for
$\tA_{ij}$, $K$ and $\tG^i$ is given by the highest spatial
derivatives, $(\partial^2\phi,\partial^2\tg,\partial \tA,\partial
K,\partial \tG)$. The evolution equations for $\tilde\gamma$ and
$\phi$ do not contain any of these highest derivatives. We define
their principal part to be given by the next highest derivatives, that
is $(\partial\phi,\partial\tg,\tilde A, K,\tG)$. Note that with this
definition of the principal part the equations are still quasilinear,
because the evolution equation for $\g_{ij}$ is linear in $K_{ij}$.
Note also that $\beta^k \partial_k$ is part of the principal part of
the equations while $\partial\beta$ terms are not.  We define the
derivative operator 
\begin{equation}
\label{defD0}
\partial_0\equiv \alpha^{-1}(\partial_t -\beta^k \partial_k).
\end{equation}
It is the derivative along the unit vector field normal to the
slices of constant time.
The principal part of the evolution equations is then
\begin{eqnarray}
\partial_0 \phi &\simeq& -{1\over 6} K, \\
\partial_0 \tg_{ij} &\simeq& -2\tA_{ij}, \\
\partial_0 K &\simeq&
-6\sigma e^{-4\phi} \tg^{ij} \phi_{,ij} , \quad {} \\
\nonumber \partial_0 \tA_{ij} &\simeq&
e^{-4\phi}\left[
- \frac{1}{2} \tg^{mn} \tg_{ij,mn} 
- 2 (1+3\sigma) \phi_{,ij}
\right. \\ \nonumber && \left.
+ a \tg_{k(i} \tG^k{}_{,j)}
+ (1-a) \tg^{kl} \tg_{k(i,j)l}
\right]^{\rm TF} \\
&& - \frac{c}{6} e^{-4\phi} \tg_{ij} \tg^{mn} \tg^{kl} \tg_{kl,mn} , \\
\label{dtGammai}
\partial_0 \tG^i &\simeq& 
2(b-1)\tg^{ij}\tg^{kl}\tA_{jk,l}-{4\over 3}b\tg^{ij}K_{,j},
\end{eqnarray}
where $\simeq$ means equal up to non-principal terms, and TF indicates
the tracefree part. We have added the constraints $aG_{(i,j)}$,
$2bM_i$ and $-(c/6)e^{-4\phi} \tg_{ij} \tg^{kl} D_{,kl}$ to the field
equations with free coefficients $a$, $b$ and $c$. Adding these terms
changes the evolution off the constraint surface which can affect the
hyperbolicity of the system. 

The principal part of the Hamiltonian
and momentum constraints is
\begin{eqnarray}
H&\equiv &H_0+e^{-4\phi}\tg^{ij}\left[a'G_{i,j}
-{c'\over 2}D_{,ij}\right], \\
H_0&\simeq
&e^{-4\phi}\tg^{ij}\left(\tg^{kl}\tg_{ki,jl}-8\phi_{,ij}\right), \\
M_i&\simeq& \tA_{ij,k}\tg^{jk}-{2\over 3}K_{,i} .
\end{eqnarray}
Here $a'$ and $c'$ parameterise different ways of writing the
Hamiltonian constraint that are found in the literature.  We shall
work explicitly only with $H_0$, and so $a'$ and $c'$ will not appear
below. There are many versions of the BSSN equations which vary in
small details in both the principal and non-principal parts.  For
comparison, the principal part of the version given in \cite{Yo2} is
characterised by $\sigma=0$ (the lapse is not densitised) and
$a=b=a'=c'=1$, $c=0$.

The constraints are compatible with the evolution equations, which means
that they form a closed evolution system. It is
\begin{eqnarray}
\partial_0 H_0 &\simeq& -2e^{-4\phi}\tg^{ij}M_{i,j}, \\
\partial_0 M_i &\simeq& \frac{1}{6} H_{0,i}
+e^{-4\phi}\Bigl(
{a\over 2}\tg^{jk}G_{i,jk} \nonumber \\
&&+{a\over 6}\tg^{jk}G_{j,ik}
+\frac{1-c}{6}\tg^{jk}D_{,ijk}\Bigr), \\
\partial_0 G_i &=& 2b M_i, \\
\partial_0 {T}&\simeq& -{c\over 2}e^{-4\phi}\tg^{ij}D_{,ij}, \\
\partial_0 D&=&-2{T}.
\end{eqnarray}


\subsection{Strong hyperbolicity of the main system}


For the purpose of decomposing 3-tensors and tensor equations, we
define $n_i$, $n^i$ and $q_{ij}$ with respect to the conformal metric
$\tg_{ij}$. In the frozen coefficients approximation, the
undifferentiated 3-metric ($\tg_{ij}$, $\tg^{ij}$ and $\phi$) is to be
treated as a background quantity, while $\tA_{ij}$, $\tG^i$, $K$,
$\tg_{ij,k}$, $\phi_{,i}$ and their derivatives are to be treated as
dynamical variables, and decomposed with respect to $n_i$.

We now prove strong hyperbolicity of the main and constraint systems
by constructing a complete set of characteristic variables $U$. 
The quantities
\begin{eqnarray}
\tilde u\equiv(\p_i\phi,\p_i\tg_{jk},\tA_{ij},\tG^i, K).
\end{eqnarray}
obey the pseudo-first order system
\begin{equation}
\partial_t \tilde u\simeq (\alpha P^i+\beta^i)\partial_i \tilde u
\end{equation}
or equivalently
\begin{equation}
\partial_0 \tilde u \simeq P^i \partial_i \tilde u.
\end{equation}
(This is not a genuine first-order system because in expressions like
$\p_i(\p_j \phi)$ that appear on its right-hand side we allow
ourselves to commute partial derivatives, rather than treating
$\p_j\phi$ as a 1-form variable $d_j$, as we would in a genuine
reduction to first order.)  As discussed in \cite{bssn1}, transversal
derivatives are automatically zero speed characteristic
variables. Here we have $P_n\p_A\phi=P_n\p_A\tg_{ij}=0$. We shall
write the eigenvalues of $P_n$ as $\lambda e^{-2\phi}$. With
this definition $\lambda$ measures the propagation speed in units of
the speed of light, measured with respect to the $\partial_0$
observers. In particular $\lambda=\pm 1$ variables propagate along the
light cone. 

In tensor components with respect to $n_i$, the scalar block of $P_n$
is given by
\begin{eqnarray}
P_n\pn\phi&=&-{1\over6} K, \\
P_n\pn\tg_{nn}&=&-2 \tA_{nn}, \\
P_n\pn\tg_{qq}&=&-2 \tA_{qq}, \\
P_n K&=&-6\sigma e^{-4\phi} \pn\phi , \\
P_n \tA_{nn}&=&e^{-4\phi}\Bigl[-\left({4\over3}+4\sigma\right)\pn\phi
+{1\over 6}\pn\tg_{nn} 
\nonumber \\ &&
+{2a\over 3}(\tG_n-\pn\tg_{nn})
\nonumber \\ &&
+\frac{1-c}{6}(\pn\tg_{nn}+\pn\tg_{qq})\Bigr], \\
P_n \tA_{qq}&=&e^{-4\phi}\Bigl[\left({4\over3}+4\sigma\right)\pn\phi
-{1\over 6}\pn\tg_{nn}
\nonumber \\ &&
-{2a\over 3}(\tG_n-\pn\tg_{nn})
\nonumber \\ &&
-\frac{1+2c}{6}(\pn\tg_{nn}+\pn\tg_{qq})\Bigr], \\
P_n\tG_n&=&-{4b\over 3} K+2(b-1) \tA_{nn}.
\end{eqnarray}
The two algebraic constraints $T=0$ and $D=0$ are often enforced in
numerical simulations after each time step. We can mimic enforcing
$T=0$ in our analysis by setting $\tA_{nn}+\tA_{qq}=0$ and
dropping one of the two variables from the system. Similarly, we can
mimic enforcing $D=0$ by setting $\p_n\tg_{nn}+\p_n\tg_{qq}=0$ and
dropping one of the two variables from the system. Both reductions
affect the hyperbolicity of the system.

\paragraph{Algebraic constraints not enforced}

The scalar block is diagonalisable for 
\begin{equation}
c>0, \quad
\eta>0, \quad
\sigma>0,\quad
c\ne \eta,
\end{equation}
with eigenvalues
\begin{equation}
\lambda=\left\{0,\pm\sqrt{c},\pm\sqrt{\eta},\pm\sqrt{\sigma}\right\},
\end{equation}
where we have defined the shorthand
\begin{equation} \label{etadef}
\eta \equiv \frac{4ab-1}{3}.
\end{equation}
(Allowing the system to be strongly hyperbolic without enforcing the
algebraic constraints is the reason why we have introduced the $c$ term,
which potentially gives $\partial_0\tA_{ij}$ a non-zero trace.)

\paragraph{Trace constraint enforced}

If we enforce $T=0$ but not $D=0$ this is consistent with the evolution
equations only for $c=0$. The scalar block is diagonalisable
for
\begin{equation}
\label{Msconditions}
c=0, \quad \eta>0,\quad \sigma>0,
\end{equation}
with 
\begin{equation}
\lambda=\left\{0,0,\pm\sqrt{\eta},\pm\sqrt{\sigma}\right\}.
\end{equation}
It would be inconsistent to enforce $D=0$ but not $T=0$.

\paragraph{Both algebraic constraints enforced}

If we enforce both algebraic constraints $D=0$ and $T=0$
the scalar block is diagonalisable for
\begin{equation}
\label{MsALconditions}
\eta>0,\quad \sigma>0.
\end{equation}
(Note that $c$ becomes irrelevant in this case.)
The characteristic variables that do not contain any transversal
derivatives are:
\begin{eqnarray}
U_0' &\equiv& (b-1) \pn\tg_{nn}+\tG_n-8b\pn\phi , \\
U'_\pm &\equiv& (1-4a) \pn\tg_{nn}+4a\tG_n -8\pn\phi \nonumber \\
&& \pm e^{2\phi}\sqrt{\eta} (6 \tA_{nn}-4{K}) , \\
V'_\pm &\equiv& 6\sqrt\sigma\pn\phi\mp e^{2\phi}{K} ,
\end{eqnarray}
with speeds 
\begin{equation}
\lambda=(0,\pm\sqrt{\eta},\pm\sqrt{\sigma}).
\end{equation}

The vector block of $P_n$ is
\begin{eqnarray}
P_n\pn\tg_{An}&=&-2\tA_{An}, \\
P_n\tA_{An}&=&e^{-4\phi}{a\over 2}\left(\tG_A-\pn\tg_{An}\right), \\
P_n\tG_A&=&2(b-1)\tA_{An},
\end{eqnarray}
as well as the trivial
$P_n\p_A\phi=P_n\p_A\tg_{nn}=P_n\p_A\tg_{qq}=0$. It is
diagonalisable for $ab>0$, with nontrivial characteristic
variables
\begin{eqnarray}
U'_A&\equiv&(b-1) \pn\tg_{nA}+\tG_A  , \\
U'_{\pm A}&\equiv& -a \pn\tg_{An}
+ a \tG_A \pm 2\sqrt{ab}e^{2\phi} \tA_{An}  ,
\end{eqnarray}
with speeds $\lambda=(0,\pm\sqrt{ab})$. 
The tensor block is
\begin{eqnarray}
P_n\pn\tg_{AB}&=&-2\tA_{AB}, \\
P_n\tA_{AB}&=&-e^{-4\phi}{1\over 2}\pn\tg_{AB},
\end{eqnarray}
as well as the trivial $P_n\p_A\tg_{Bn}=0$.
It is always diagonalisable, with nontrivial characteristic variables
\begin{equation}
U'_{\pm AB}\equiv {1\over 2} \pn\tg_{AB} \mp e^{2\phi}  \tA_{AB}  ,
\end{equation}
with speeds $\lambda=\pm 1$.
We see that the vector and tensor sectors do not add any new
conditions for strong hyperbolicity.

In summary, we have shown that the BSSN system with both algebraic
constraints enforced continuously is strongly hyperbolic if $\sigma>0$
and $\eta>0$. If neither algebraic constraint is enforced, $c>0$ and
$c\ne \eta$ are also required. On the other hand, if only the trace
constraint is enforced, $c=0$ is required. (If both constraints are
enforced, $c$ is irrelevant and can be set to zero.)  The
characteristic spectrum of the complete system is
\begin{equation}
\lambda=\{0,\pm\sqrt{\sigma},\pm\sqrt{\eta},\pm\sqrt{ab},\pm 1\} .
\end{equation}


\subsection{Strong hyperbolicity of the constraint system}


We construct characteristic variables for the constraint
system from the set
\begin{equation}
\tilde u_c\equiv(H_0, M_i, \p_i G_j, \p_i T, \p_i\p_j D).
\end{equation}

The scalar sector of the constraint propagation is
\begin{eqnarray}
P_n H_0&=& -2 e^{-4\phi} M_n, \\
P_n M_n&=& \frac{1}{6} H_0
+e^{-4\phi}\left({2a\over3} \pn G_n+\frac{1-c}{6} \pn^2 D \right), \\
P_n \pn G_n&=&2b M_n, \\
P_n \pn T&=&-{c\over 2}e^{-4\phi} \pn^2 D , \\
P_n \pn^2 D &=&-2 \pn T.
\end{eqnarray}
If we define
\begin{equation}
C'_\pm\equiv 4a \pn G_n \pm 6 \sqrt\eta e^{2\phi} M_n 
+ e^{4\phi} H_0
\end{equation}
the characteristic variables are
\begin{eqnarray}
C_0' \equiv e^{4\phi}b H_0+ \pn G_n ,  \\
\sqrt{c}\, \pn^2 D\mp2e^{2\phi} \pn T  , \\
(1-c)\left( \pn^2 D\mp {2\over \sqrt{\eta}} 
e^{2\phi} \pn T\right) +\left(1-{c\over\eta}\right) C'_\pm ,
\end{eqnarray}
with speeds $\lambda=(0,\pm\sqrt{c},\pm\sqrt{\eta})$.
If we optionally impose the trace constraint, the characteristic
variables are $ C'_0$ and $ C'_\pm + \pn^2 D$
with speeds $\lambda=(0,\pm\sqrt{\eta})$.
If we impose both the trace and determinant constraints, we are left
with $ C'_0$ and $ C'_\pm$ with speeds $\lambda=(0,\pm\sqrt{\eta})$.

The vector sector of the constraint propagation is
\begin{eqnarray}
P_n M_A&=&e^{-4\phi}{a\over 2} \pn G_A, \\
P_n \pn G_A&=&2b M_A,
\end{eqnarray}
as well as $P_n\p_A G_n=P_n\p_A T=P_n\p_A\pn D=0$.
The nontrivial characteristic variables are
\begin{equation}
C'_{\pm A}\equiv a \pn G_A\pm 2\sqrt{ab}e^{2\phi} M_A  .
\end{equation}
with speeds $\lambda=\pm\sqrt{ab}$.
The tensor sector is completely trivial, with $P_n\p_A
G_B=P_n\p_A\p_B T=0$ (including the traces).
The constraint system, in all three cases, is strongly hyperbolic as
long as the main system is strongly hyperbolic.


\subsection{Main system energy}


We look for an energy density $\epsilon$ of the second-order system
that is positive definite in $\tA_{ij}$, $K$, $\tG_i$, $\tg_{ij,k}$
and $\phi_{,i}$ and obeys a conservation law.  With the shorthands
$t_i\equiv \tg^{jk}\tg_{jk,i}=D_{,i}$ and
$d_i\equiv\tg^{jk}\tg_{ij,k}$, the most general quadratic form in
these variables is 
\begin{eqnarray}
\epsilon 
&=& 
  c_0 e^{4\phi} \, \tA_{ij}{}^2
+ c_1 \, \tG_i{}^2
+ c_2 \, t_i{}^2
+ c_3 \, d_i{}^2
\nonumber \\
&&+
  c_4 \, d_i \tG^i
+ c_5 \, t_i \tG^i
+ c_6 \, d_i t^i
+ c_7 e^{4\phi} \, K^2
\nonumber \\
&&+
  c_8 \, \tg_{ij,k}{}^2
+ c_9 \, \tg_{ij,k}\tg^{ik,j}
+ c_{10} \, \tG_i\phi^{,i}
+ c_{11} e^{4\phi} \, TK 
\nonumber \\
&&+
  c_{12} \, \phi_{,i}{}^2
+ c_{13} \, d_i\phi^{,i}
+ c_{14} \, t_i\phi^{,i}
+ c_{15} e^{4\phi} \, T^2 .
\end{eqnarray}
In the cross term $\tg_{ij,k}\tg^{ik,j}$ and similar terms in the
remainder of the paper indices are raised only after differentiation.
The general ansatz for the flux $F^i$ contains 14 free coefficients.
After some linear algebra it is possible to see that this $\epsilon$
and $F^i$ together obey a conservation law only if
\begin{equation}
\label{BSSNsigmacondition}
4ab=1+3\sigma,
\end{equation}
or equivalently $\eta=\sigma$.

There are additional restrictions depending on whether or not we
enforce the algebraic constraints during evolution. If we work with
non-vanishing $D$ and $T$ then we need $ab=\sigma=1$ and
$c_{11}(c-1)=0$. On the other hand, if we enforce both algebraic
constraints the parameter $c$ and several of the coefficients $c_k$
become irrelevant, with no additional constraints on the parameters.

From now on, in order to simplify the equations, we enforce both the
$T=0$ and $D=0$ constraints in the remainder of this Section.

The generic conserved energy depends on four free coefficients, one of
which is an overall factor. With the shorthands
\begin{eqnarray}
\epsilon_0 &=& e^{4\phi} \tA_{ij}{}^2+\frac{1}{4} \tg_{ij,k}{}^2+
        \frac{2a-ab-1}{2}\tg_{ij,k}\tg^{ik,j} \nonumber \\
&&- ad_i(\tG^i-8b\phi^{,i}), \\
\epsilon_1 &=& \left[\tG_i-8b\phi_{,i}+(b-1)d_i\right]^2 , \\
\epsilon_2 &=& d_i{}^2-\tg_{ij,k}\tg^{ki,j} , \\
\epsilon_3 &=& e^{4\phi} K^2+36\sigma \phi^{,i}\phi_{,i}
\end{eqnarray}
the energy density is
\begin{equation}
\epsilon =c_0\epsilon_0 + c_1\epsilon_1+ [c_3-(b-1)^2c_1] \epsilon_2 
+ c_7 \epsilon_3 .
\end{equation}
The free parameters $c_0$, $c_1$, $c_3$ and $c_7$  will be
restricted below by inequalities derived from the requirement that
$\epsilon$ be positive definite.

This energy reduces to that given in \cite{SarbachBSSN} for a
first-order reduction of BSSN with the change of notation
$\sigma\to 2\sigma$, $b\to m$, $\tg_{ij,k}\to d_{kij}$ and
$\phi_{,i}\to d_i/12$, the restriction
$a=1$ to the parameters of the evolution equations, fixing the overall
scale as $c_0=1$, and the further
restrictions $c_1=1/(2b-2)$, $c_3=0$ and
$c_7=1/\sigma=3/(4b-1)$ on the coefficients
of the energy. The condition (\ref{BSSNsigmacondition}) on $\sigma$
reduces to a similar condition in \cite{SarbachBSSN}. The choice of
Sarbach et al. can be interpreted as follows: $c_3=0$ and
$c_1=a/(2b-2)$ eliminate $d_i$ from the energy
($c_3=c_4=c_{13}=0$); assuming those conditions, $a=1$ is the
only choice that eliminates the cross term $\tg_{ij,k}\tg^{ki,j}$.
Their choice for $c_7$ has no relevant effect. Note that the 
positivity condition on their choice of 
$c_1$ forces $b>1$, and together with $a=1$
this gives rise to a superluminal speed. By contrast, by retaining
the contractions $t_i$ and $d_i$ in the ansatz for the energy we shall be
able to make all speeds physical.

The four terms in the energy are conserved separately:
\begin{eqnarray}
\partial_0 \epsilon_0 &=& \partial_i F_0^i , \\
\partial_0 \epsilon_1 &=& 0 , \\
\partial_0 \epsilon_2 &=& \partial_i F_2^i , \\
\partial_0 \epsilon_3 &=& \partial_i F_3^i.
\end{eqnarray}
(Condition (\ref{BSSNsigmacondition}) is 
required only for the conservation of $\epsilon_0$.)
The fluxes are given by
\begin{eqnarray}
F_0^i &=&
2\tA^{ij}[a\tG_j-a(b-1)d_j-2(1+3\sigma)\phi_{,j}] \nonumber
\\
&& + \tg^{il}\tA^{jk}[ 2(1+ab-2a)\tg_{jl,k} - \tg_{jk,l} ] , \\
F_2^i &=& 4 \tA^{jk} (\tg^{im}\tg_{mj}{}_{,k}-\tg^i{}_j\tg_{km}{}^{,m}),
\\
F_3^i &=& -12\sigma K \phi^{,i} .
\end{eqnarray}
The energy density in terms of characteristic variables is
\begin{eqnarray}
\epsilon &=&
\frac{c_0}{2}(U_{+AB}^2+U_{-AB}^2)
+\frac{c_0}{4ab}(U_{+A}^2+U_{-A}^2) \nonumber \\
&&+\frac{c_0 c_7}{16\sigma(2c_0+3c_7)} (U_+^2+U_-^2)
+\frac{2c_0+3c_7}{6} (V_+^2+V_-^2)  \nonumber \\
&& + \hbox{ quadratic in zero speed variables},
\end{eqnarray}
and the flux is
\begin{eqnarray}
e^{2\phi}F^n&=&
\frac{c_0}{2}(U_{+AB}^2-U_{-AB}^2) 
+\frac{c_0}{4\sqrt{ab}}(U_{+A}^2-U_{-A}^2)\nonumber \\
&&+\frac{c_0c_7}{16\sqrt{\sigma}(2c_0+3c_7)} (U_+^2-U_-^2) \nonumber\\
&&+\frac{2c_0+3c_7}{6}\sqrt{\sigma}\left(V_+^2-V_-^2\right)
\end{eqnarray}
We have modified the characteristic variables by adding terms in
$\partial_A\tg_{ij}$ in order to write $\epsilon$ in terms of
characteristic variables. With the shorthands
\begin{eqnarray}
z_1 &=& \frac{c_3-(b-1)^2c_1}{c_0} , \\
z_2 &=& 1+3ab-4a+6z_1 , \\
z_3 &=& a-ab-2z_1 , \\
z_4 &=& 2+3ab-5a+6z_1 , \\
z_5 &=& 1+ab-2a+2z_1 , \\
z_6 &=& \frac{c_0}{2\sqrt\sigma(2c_0+3c_7)} ,
\end{eqnarray}
the modified characteristic variables are
\begin{eqnarray}
\label{BSSN_full_Upm}
U_\pm &=& (1-4a)\partial_ng_{nn} +4a\tG_n - 8\partial_n\phi
\nonumber \\ &&
\pm \sqrt\sigma e^{2\phi}(6\tA_{nn}-4K) -2z_2\partial^Ag_{An} , \\
V_\pm &=& 6\sqrt\sigma \partial_n\phi\mp e^{2\phi} K 
- z_6 U_\pm , \\
U_{\pm A} &=& \pm 2\sqrt{ab}e^{2\phi}\tA_{An}+a\tG_A 
       - a\partial_n g_{An} \nonumber \\
    && +z_3\partial^Bg_{AB}+\frac{z_4}{2}\partial_Ag_{nn}-8ab\partial_A\phi , \\
U_{\pm AB} &=& \mp e^{2\phi} \tA_{AB} + \frac{1}{2} \partial_n g_{AB}
               \nonumber \\
           && - \frac{z_5}{2}(\partial_A g_{Bn}+\partial_B g_{An}
	        - q_{AB}\partial^C g_{Cn}) .
\end{eqnarray}
The expressions for $U_0$ and $U_A$ are not given, but they also
include transversal derivatives. 
Note that because $\eta=\sigma$, $U'_\pm$ and $V'_\pm$ are
eigenvectors for the same eigenvalues, and we have used this to make
the energy diagonal in $U_\pm$ and $V_\pm$.


\subsection{Constraint system energy}


The constraint energy is quadratic in $H_0$, $M_i$ and $G_{i,j}$.
(We must use $G_{i,j}$ rather than $G_i$ so that all terms are of
the same order in derivatives). The most general conserved expression is
\begin{eqnarray}
\epsilon_c&=&
w_0 e^{8\phi} H_0^2
+12 w_1 e^{4\phi} M_i^2
+\frac{3a w_1}{b}(G_{i,j})^2
\nonumber \\ 
&&+ w_2 G_{i,j}G^{j,i}
+2\frac{w_0+w_1}{b}e^{4\phi}{G_i}^{,i}H_0
\nonumber \\
&&+ \frac{w_0+(1+ab)w_1-b^2 w_2}{b^2} ({G_i}^{,i})^2 ,
\end{eqnarray}
with arbitrary coefficients $w_0$, $w_1$ and $w_2$. The corresponding flux
is
\begin{eqnarray}
F_c^i &=& 4w_1 ( e^{4\phi}H_0M^i+3aG^{j,i}M_{j}+aM^iG^j{}_{,j})
\nonumber \\
&&+4b w_2 (G^{i,j}M_j-M^iG^j{}_{,j}) .
\end{eqnarray}

In terms of characteristic variables, the constraint energy and flux
are
\begin{eqnarray}
\epsilon_c &=& 
\frac{3w_1}{2ab}(C_{-A}C_-^A+C_{+A}C_+^A)
+\frac{w_1}{6\sigma}(C_-^2+C_+^2)
\nonumber \\
&&+ \hbox{ quadratic in zero speed variables}
\label{epsilonc_bssn}, \\
e^{2\phi} F^n_c &=& 
\frac{3w_1}{2\sqrt{ab}}(C_{+A}C_+^A-C_{-A}C_-^A) \nonumber \\
&& \qquad + \frac{w_1}{6\sqrt\sigma}(C_+^2-C_-^2), 
\label{fluxc_bssn}
\end{eqnarray}
where the modified non-zero speed  characteristic variables are
\begin{eqnarray}
C_\pm &=& e^{4\phi} H_0 \pm 6 e^{2\phi} \sqrt\sigma M_n 
          +4a\partial_n G_n \nonumber \\
&& + \frac{aw_1-bw_2}{w_1} \partial^AG_A , \\
C_{\pm A} &=& \pm 2\sqrt{ab} e^{2\phi} M_A + a\partial_n G_A
          +\frac{w_2 b}{3 w_1}\partial_A G_n .
\end{eqnarray}
Note that $w_2$ does not appear explicitly in
(\ref{epsilonc_bssn}) or (\ref{fluxc_bssn}), but it does appear in the
definition of the characteristic variables.


\subsection{Symmetric hyperbolicity and causal speeds}


With the condition $\eta=\sigma$ that is required for conservation of
the main energy, the complete spectrum can be written entirely in
terms of $\sigma$:
\begin{equation}
\lambda=\left(0,\pm\sqrt{\sigma},\pm{\sqrt{1+3\sigma}\over 2},\pm
1\right).
\end{equation}
Strong hyperbolicity requires $\sigma>0$ and the absence of
superluminal speeds requires $\sigma\le 1$. Note that with
$\sigma=ab=1$ all speeds are either 0 or $\pm1$. 

Now we impose positive definiteness of the energies, that is,
all eigenvalues of their respective matrices are strictly
positive. This means that the main energy vanishes if and only if 
$\tg_{ij}$ and $\phi$ are constant and all other variables vanish, and
that the constraint energy vanishes if and only if $G_i$ is constant
and the Hamiltonian and momentum constraints vanish. (We are assuming
that the two algebraic constraints vanish because they are being
continuously re-imposed.)

To see when the constraint energy is positive definite, we need
to decompose $G_{i,j}$ as
\begin{equation}
\label{Gijdecomposition}
G_{i,j}=S_{ij}+A_{ij}+{1\over 3}\gamma_{ij}{G^k}_{,k}
\end{equation}
where $S_{ij}$ is symmetric and tracefree and $A_{ij}$ is
antisymmetric. We obtain a sum of simple squares and a quadratic form
in the two variables $H_0$ and ${G^i}_{,i}$.

The constraint energy is positive definite if and only if
\begin{eqnarray}
&& w_0>0, \quad 0 < w_1 < 3 w_0, \quad
\frac{w_0+w_1}{4w_0} < ab \le 1, \nonumber \\
&&-ab < \frac{w_2b^2}{3w_1} < ab - \frac{w_0+w_1}{2w_0} .
\end{eqnarray}
It is interesting to see how $w_1$ affects the possible range for $ab$.

The positivity of the main energy is more complicated. We must
take into account that the partial traces $t_i$ and $d_i$ of the
three-index object $\tg_{ij,k}$ appear both explicitly and explicitly
in the energy. We therefore decompose it as
\begin{equation}
\tg_{ij,k}={1\over
  5}\left[(3d_{(i}-t_{(i})\tg_{j)k}+(2t_k-d_k)\tg_{ij}\right]
+f_{ijk}
\end{equation}
where $f_{ijk}$ is completely tracefree. We then obtain a quadratic
form in $t_i$ and $d_i$, plus $c_8 f_{ijk}^2+c_9 f_{ijk}f^{ikj}$. To
analyse the positivity of the latter we decompose $f_{ijk}$ in its 12
independent frame components and analyse the corresponding quadratic
form by brute force. It turns out to be positive if and only if
$c_8>0$ and $-c_8<c_9<2c_8$.  Finally, assuming causal speeds, we
obtain the positivity conditions
\begin{eqnarray}
&&c_0>0 , \qquad
c_7>0 , \qquad
2a^2 c_0 < (4ab-1) c_1 , \nonumber\\
&&5c_0\frac{2a^2c_0-(4ab-1)c_1}{36c_1} \nonumber \\
&&<c_3-(b-1)^2c_1 
+\frac{1+2ab-4a}{4}c_0 <0 . 
\end{eqnarray}

The three sets of inequalities are compatible, and can be solved
sequentially, in this order: first choose $c_0$, $c_7$, $w_0$ and $w_1$
independently, then $a$ and $b$, followed by $c_1$.  Finally choose
$c_3$ and $w_2$. It is clear from this construction that the solutions
form a single connected set. For example, a possible solution with
$\lambda=(0,\pm 1)$ is
\begin{eqnarray}
&&a=b=\sigma=1, \label{BSSNexample1}\\
&& c_0=c_7=1, \quad c_1=2, \quad c_3=0, \label{BSSNexample2}\\
&&w_0=1, \quad w_1={1\over 2} , \quad w_2=0. \label{BSSNexample3}
\end{eqnarray}


\section{The NOR formulation of the Einstein equations}
\label{section:nor}



\subsection{Field equations}


In this Section, all indices are moved with $\gamma_{ij}$ and
its inverse $\gamma^{ij}$. The NOR system is obtained from the ADM
system with densitised lapse by introducing the variables
\begin{equation}
f_i\equiv{\gamma_{ij}}^{,j}-{\rho\over2}\gamma^{jk}\gamma_{jk,i},
\end{equation}
where $\rho$ is a constant parameter. In \cite{NOR} the choice
$\rho=1$ is made. $\rho=2/3$ makes $f_i=\tg_{ij}\tG^j$. 
The definition of $f_i$ gives rise to the constraint
\begin{equation}
G_i\equiv f_i-{\gamma_{ij}}^{,j}+{\rho\over2}\gamma^{jk}\gamma_{jk,i}=0.
\end{equation}
As in the BSSN system, we parameterise the use of this
constraint with $a$ and the use of the momentum constraint with $b$ by
adding $aG_{(i,j)}$ and $2bM_i$ to the evolution equations for $K_{ij}$
and $f_i$. Following \cite{NOR}, we also add
$c\gamma_{ij}H$ to the evolution equation for $K_{ij}$ with free
parameter $c$.  The principal part of the evolution equations is
\begin{eqnarray}
\partial_0 \g_{ij}&\simeq&-2K_{ij}, \\
\partial_0 K_{ij}&\simeq&{1\over2}\Bigl[a(f_{i,j}+f_{j,i})-{\g_{ij,k}}^{,k}
\nonumber \\ &&
+(1-a)({\g_{ki,j}}^{,k}+{\g_{kj,i}}^{,k})
\nonumber \\ &&
+(a\rho-1-\sigma)\g^{kl}\g_{kl,ij}\Bigr] +c \g_{ij}H, \\
\partial_0 f_i&\simeq&2(b-1){K_{ij}}^{,j}+(\rho-2b)K_{,i}, 
\end{eqnarray}
where
\begin{eqnarray}
H&\equiv&H_0+a'{G_i}^{,i}, \\
H_0&\simeq& {\g_{ij}}^{,ij}-\g^{ij}{\g_{ij,k}}^{,k}, \\
M_i&\simeq& {K_{ij}}^{,j}-K_{,i}.
\end{eqnarray}
In \cite{NOR}, $\rho=a=a'=1$. 

The constraints system is
\begin{eqnarray}
\partial_0 H_0 &\simeq& -2M{_i}^{,i}, \\
\partial_0 M_i &\simeq&-\left({1\over 2}+2c\right)H_{0,i} 
\nonumber \\ && 
+{a\over 2} {G_{i,j}}^{,j}-\left({a\over 2}+2ca'\right){G_{j,i}}^{,j},
\\
\partial_0 G_i&=& 2 b M_i .
\end{eqnarray}


\subsection{Strong hyperbolicity}


This proceeds exactly as in the BSSN system, with pseudo-first
order variables
\begin{equation}
\tilde u\equiv(\p_i\g_{jk}, K_{ij}, f_i)
\end{equation}
and
\begin{equation}
\tilde u_c=\{ H_0, M_i, \p_i G_j\}
\end{equation}
for the constraint system. We define $n_i$, $n^i$ and $q_{ij}$
with respect to $\g_{ij}$.  (In \cite{NOR}, a
flat auxiliary metric is introduced instead.)

The scalar sector of the main system is diagonalisable for $\chi>0$,
$\sigma>0$ and $\chi\ne\sigma$, with
\begin{equation}
\lambda=\{0,\pm\sqrt{\chi},\pm\sqrt{\sigma}\}, 
\end{equation}
with the shorthand
\begin{equation}
\chi\equiv 1+4c(1-a'b). 
\end{equation}
If $\eta=\chi$ we can diagonalise the scalar sector also for
$\chi=\sigma$. (Note that while we use the same shorthand $\eta$
defined by (\ref{etadef}) as in the BSSN system, $\sqrt\eta$ is not a
speed in the NOR system.)

The vector sector is diagonalisable for $ab>0$ with 
\begin{equation}
\lambda=\{0,\pm\sqrt{ab}\},
\end{equation}
and the tensor sector is always diagonalisable, with $\lambda=\pm 1$. 
For general values of $(\rho,\sigma,a,b,c,a')$ the characteristic
variables are too long to give here.

The scalar sector of the constraint system is
diagonalisable for $\chi>0$ with characteristic variables
\begin{eqnarray}
C_0\equiv  \pn G_n+b H_0 ,\\
C_\pm \equiv (1+4c) H_0+4ca'  \pn G_n\mp 2\sqrt{\chi} M_n ,
\end{eqnarray}
with speeds $\lambda=\{0,\pm\sqrt\chi\}$,
and the vector sector is diagonalisable for $ab>0$ with 
\begin{equation}
C_{\pm A}\equiv  a \pn G_A\pm 2\sqrt{ab} M_A ,
\end{equation}
with speeds $\lambda=\{\pm\sqrt{ab}\}$.

The union of conditions for both the main system and the
constraint system to be strongly hyperbolic, and for all speeds to be
physical ($|\lambda|\le 1$), is
\begin{equation}
\label{physical}
0<\sigma\le 1, \quad
0<ab\le 1, \quad 
0<\chi\le 1,
\end{equation}
and either $\chi\not=\sigma$, or $\chi=\sigma=\eta$. The union of all
speeds is
\begin{equation}
\lambda=\{0,\pm 1,\pm\sqrt{ab}, \pm \sqrt{\sigma},\pm \sqrt{\chi}\}.
\end{equation}
We see that we can make all of these speeds either zero or one
by choosing
\begin{equation}
ab=\sigma=\chi=1 \Rightarrow c(1-a'b)=0 ,
\end{equation}
that is, either $c=0$ or $a'=a$. From now on we restrict ourselves to
the choice $c=0$. This makes $\chi=1$, and removes $a'$ from the
system. We keep $a$, $b$, $\rho$ and $\sigma$ free, with the single
restriction that $\sigma=1$ requires $ab=1$. The value of $\rho$ is
irrelevant for strong hyperbolicity. Our results on strong
hyperbolicity of the NOR system confirm and generalise those of
\cite{NOR}, which were obtained using a pseudo-differential reduction
to first order.


\subsection{Main system energy}


The most general ansatz for the energy density $\epsilon$ is quadratic
in $\gamma_{ij,k}$, $K_{ij}$ and $f_i$. As well as contracting the
free indices on pairs of these, we can form the contractions $K$,
$d_i\equiv{\g_{ij}}^{,j}$ and $t_i\equiv\g^{jk}\g_{jk,i}$ first:
\begin{eqnarray}
\epsilon 
&=& 
  c_0 \, K_{ij}{}^2
+ c_1 \, f_i{}^2
+ c_2 \, t_i{}^2
+ c_3 \, d_i{}^2
\nonumber \\
&&+
  c_4 \, d_i f^i
+ c_5 \, t_i f^i
+ c_6 \, d_i t^i
+ c_7 \, K^2
\nonumber \\
&&+
  c_8 \, \gamma_{ij,k}{}^2
+ c_9 \, \gamma_{ij,k}\gamma^{ik,j}.
\end{eqnarray}
The most general form that is conserved depends on four free
parameters $c_0$, $c_1$, $c_3$ and $c_7$ obeying
\begin{equation}
2(ab-1)c_7 = (\sigma-2ab+1)c_0 .
\end{equation}
for arbitrary $(\rho,\sigma,a,b)$. The coefficient $c_0$ must be
strictly positive and therefore there are two possibilities: either
$ab=1$ (which now implies $\sigma=1$) and then $c_0$ and $c_7$ are
independent, or $ab\not=1$ and then $c_7$ is determined by $c_0$ and
the parameters $(\sigma,a,b)$. What follows is valid in both cases,
with the corresponding restrictions. Note that strong hyperbolicity
and energy conservation together give the two directions of
$\sigma=1\Leftrightarrow ab=1$.

With the shorthands
\begin{eqnarray}
\epsilon_0 &=& K_{ij}{}^2
+\frac{1}{4}\gamma_{ij,k}{}^2
+\frac{2a-ab-1}{2}\gamma_{ij,k}\gamma^{ik,j}
\nonumber \\ &&
-ad_i\left[f^i -\left(b-{\rho\over2}\right)t^i\right] ,
\\
\epsilon_1 &=& \left[ f_i -\left(b-{\rho\over 2}\right)t_i+(b-1)d_i\right]^2 ,
\\
\epsilon_2 &=& d_i{}^2-\gamma_{ij,k}\gamma^{ik,j} ,
\\
\epsilon_3 &=& K^2+\frac{2+2ab-2a\rho+\sigma}{4}t_i^2 \nonumber \\
&& +(a-1)d_it^i-a t_i f^i 
\end{eqnarray}
(the indices on $\g_{ij,k}$ are raised only after differentiation),
the energy can be written as
\begin{equation}
\epsilon =  c_0 \epsilon_0 + c_1 \epsilon_1 
+ [c_3-(b-1)^2c_1] \epsilon_2 + c_7 \epsilon_3 .
\end{equation}
The flux is
\begin{equation}
F^i = c_0 F_0^i + [c_3-(b-1)^2c_1] F_2^i +c_7 F_3^i,
\end{equation}
with
\begin{eqnarray}
F_0^i &=& 2a K^{ij} \left[ f_j-(b-1)d_j-(b-\rho/2)t_j\right]\nonumber\\
&&+ \gamma^{il}K^{jk}\left[2(1+ab-2a)\gamma_{lj,k}-\gamma_{jk,l}\right]
,\\
F_2^i &=& 4 \gamma^{il}K^{jk} (\gamma_{lj,k}-\gamma_{lj}d_k) , \\
F_3^i &=& K \left[2af^i + (a\rho-2-\sigma)t^i+2(1-a)d^i\right] \nonumber
\\
&&+ 2(1-ab)K^{ij}t_j .
\end{eqnarray}
(As in BSSN, $\epsilon_1$ has no flux.) The flux in terms of
characteristic variables is
\begin{eqnarray}
F^n&=&
\frac{c_0}{2}(U_{+AB}^2-U_{-AB}^2)
+\frac{c_0}{4\sqrt{ab}}(U_{+A}^2-U_{-A}^2)
\nonumber \\
&&+\frac{c_0+c_7}{2\sqrt{\sigma}}(V_+^2-V_-^2)
+\frac{c_0}{16}\,\frac{c_0+3c_7}{c_0+c_7}(U_+^2-U_-^2)
, \nonumber\\
\end{eqnarray}
The characteristic variables
including transversal derivatives are
\begin{eqnarray}
U_\pm&=&-\partial_n \g_{qq}-2z_2\partial^A\g_{An}\pm 2K_{qq}, \\
V_\pm&=&af_n
\pm \sqrt\sigma K_{nn} 
+\frac{a\rho-\sigma-1}{2}\partial_n\g_{qq}
\nonumber \\
&&+\frac{a\rho-\sigma-2a}{2} \partial_n \g_{nn}
+ z_3 \partial^A\g_{An}
\label{VpmNOR} 
\\
&&+\frac{c_7}{2(c_0+c_7)}(
-\partial_n \g_{qq}-2z_2\partial^A\g_{An}\pm 2\sqrt\sigma K_{qq}) ,
\nonumber \\
U_{\pm A}&=&af_A-a\partial_n\g_{An} \pm 2\sqrt{ab}K_{An} \nonumber\\
&& +z_3\partial^B\g_{AB} +\partial_A (z_4\g_{nn}+z_5\g_{qq})
, \\
U_{\pm AB}&=&\mp K_{AB}+\frac{1}{2}\partial_n\g_{AB}
\nonumber \\
&&+z_2\left(\partial_{(A}\g_{B)n}-\frac{1}{2}q_{AB}\p^C\g_{Cn}\right),
\end{eqnarray}
with speeds $\lambda=\{\pm 1, \pm\sqrt{\sigma},\pm\sqrt{ab},\pm 1\}$,
where we have defined the shorthands
\begin{eqnarray}
z_1 &=& \frac{c_3-(b-1)^2c_1}{c_0} , \\
z_2 &=& 2a-ab-1-2z_1 , \\
z_3 &=& a-ab-2z_1 , \\
z_4 &=& (2ab-\sigma-1)/2+1-2a+a\rho/2+2z_1 , \\
z_5 &=& (2ab-\sigma-1)/2+a(1-3b+\rho)/2-z_1 .
\end{eqnarray}
Note that the last term of (\ref{VpmNOR}) requires $c_0+c_7\not=0$.
We shall see below that this is always true.


\subsection{Constraint system energy}


The constraint energy is quadratic in $H_0$, $M_i$ and $G_{i,j}$.
The most general form that is conserved has three free parameters
$w_0$, $w_1$ and $w_2$
for arbitrary $(\rho,\sigma,a,b)$. It is
\begin{eqnarray}
\epsilon_c&=&w_0H_0^2+4w_1M_i^2 +{aw_1\over b}(G_{i,j})^2
\nonumber \\ &&
+w_2G_{i,j}G^{j,i} +2{w_0-w_1\over b}{G^i}_{,i}H_0 
\nonumber \\ &&
+{w_0-(1+ab)w_1 -b^2 w_2\over b^2}({G^i}_{,i})^2
\end{eqnarray}
The flux is
\begin{eqnarray}
F^i_c&=&-4(aw_1+bw_2){G^j}_{,j}M^i-4w_1H_0M^i
\nonumber \\ &&
+4(bw_2 G^{i,j}+aw_1G^{j,i})M_j.
\end{eqnarray}
In terms of characteristic variables,
\begin{equation}
F^n_c={w_1\over 2}(C_+^2-C_-^2)
+{w_1\over 2\sqrt{ab}}(C_{+A}^2-C_{-A}^2),
\end{equation}
where, including transversal derivatives,
\begin{eqnarray}
C_\pm&=&H_0+\left(a+{bw_2\over w_1}\right)\partial_AG^A
\mp 2M_n, \\
C_{\pm A}&=&a\partial_n G_A+{bw_2\over w_1}\partial_A G_n
\pm2\sqrt{ab}M_A,
\end{eqnarray}
with speeds $\lambda=\{\pm 1, \pm\sqrt{ab}\}$.


\subsection{Symmetric hyperbolicity and causal speeds}


Positivity of the constraint energy requires that
for a given value of $ab$ we must choose $w_0$, $w_1$ and $w_2$ obeying
\begin{eqnarray}
w_0 >0 , \quad 0<\frac{w_1}{w_0}<1, \quad 
b^2\frac{w_2}{w_1}<\frac{3}{4}\left(1-\frac{w_1}{w_0}\right) ,
\nonumber \\
b^2\frac{|w_2|}{w_1} < ab < \frac{3}{2}\left(1-\frac{w_1}{w_0}\right)
-b^2\frac{w_2}{w_1} . \quad
\end{eqnarray}
The positivity conditions of the main energy will be analysed
separately for the generic and special cases.

In the general case $ab\ne 1$ and $\sigma\ne 1$ positivity of the main
energy requires that (using the strong hyperbolicity condition
$\sigma>0$ and the causal speeds condition $\sigma\le 1$)
\begin{eqnarray}
&&c_0>0, \quad \frac{1+3\sigma}{4}< ab < 1 , \nonumber \\
&&a^2 c_0(1-\sigma) < 2\sigma c_1(1-ab), \nonumber\\
&&-\frac{5c_0}{4}\,\frac{2\sigma c_1(1-ab)-a^2c_0(1-\sigma)}{2c_1(1-ab)(4-4ab+3\sigma)-a^2c_0(4ab-1-3\sigma)} 
\nonumber \\
&&<  c_3-(b-1)^2c_1 +\frac{1+2ab-4a}{4}c_0 < 0 . 
\end{eqnarray}
The value of $\rho$ is irrelevant also for symmetric
hyperbolicity. The restrictions on $\sigma$ and $ab$ guarantee that $c_0+c_7$
and $c_0+3c_7$ are always strictly positive. A simple example is
\begin{eqnarray}
&&a=\frac{3}{4}, \quad b=1, \quad \sigma=\frac{1}{2}, 
\quad \rho=\frac{2}{3} ,  \\
&&c_0=1, \quad c_1=2, \quad c_3=c_7=0, \\
&&w_0=1, \quad w_1={1\over 4} ,\quad w_2=0.
\end{eqnarray}

In the special case $ab=\sigma=1$ positivity of the main energy requires
\begin{eqnarray}
&&c_0>0, \quad c_0+3c_7>0, \quad b^2c_1 > c_0+c_7>0, \nonumber \\
&&-\frac{5c_0}{4} \, \frac{b^2c_1-c_0-c_7}{3b^2c_1-c_0-3c_7} \nonumber
\\ &&<
c_3-(b-1)^2c_1+\frac{3b-4}{4b}c_0 < 0 .
\end{eqnarray}
A simple example is 
\begin{eqnarray}
&&a=b=\sigma=1, \quad \rho=\frac{2}{3}, \label{NORexample1} \\
&&c_0=1, \quad  c_1=3, \quad c_3=c_7=0, \label{NORexample2} \\
&&w_0=1, \quad  w_1={1\over 4}, \quad w_2=0. \label{NORexample3}
\end{eqnarray}


\section{Constraint-preserving boundary conditions}
\label{section:boundary}



\subsection{The boundary system}


We have proved symmetric hyperbolicity for both the main system and
the constraint system of the BSSN and NOR formulations. Therefore both
formulations admit constraint-preserving boundary conditions of the
type proposed in \cite{CalabreseCPBC}. A general discussion is left to
future work. Here we summarize the basic idea, and propose a family of
boundary conditions for the case of a smooth boundary with tangential
shift.

Because the constraint system is compatible with the evolution system,
for every pair of characteristic variables $C_\pm$ of the constraint
system with speeds $\pm\lambda$, there is a pair of characteristic
variables $U_\pm$ with the same speeds, such that (after suitable
normalisation), they obey
\begin{equation}
C_\pm=\pn U_\pm + \dots,
\end{equation}
where the dots, here and in the remainder of this subsection, indicate
transversal derivatives and lower-order terms. In order to guarantee
that the constraint energy does not grow, we formally impose the
homogeneous maximally dissipative boundary condition
\begin{equation}
\label{CPBC}
C_+-\kappa C_-=0
\end{equation}
on the constraint system. This must then be translated into a boundary
condition on the main system. From
(\ref{defD0}), we have
\begin{equation}
\p_t U_\pm=(\pm\lambda\alpha+\beta_n)\pn U_\pm + \dots
\end{equation}
In the following we restrict consideration to the case where
$\beta_n=0$ on the boundary, so that the $\lambda=0$ characteristic
variables propagate along the boundary. (\ref{CPBC}) is then equivalent to
\begin{equation}
\label{dotform}
\p_tU_++\kappa\p_tU_-=\dots
\end{equation}
We define a variable $X\equiv U_+ +\kappa U_-$ that is restricted to
the boundary, with evolution equation
\begin{equation}
\label{Xdot}
\p_t X=\dots, 
\end{equation}
and impose the boundary condition
\begin{equation}
U_++\kappa U_-=X
\end{equation}
on the main system. 

We have chosen our notation in the previous sections so that in both
BSSN and NOR we have
\begin{eqnarray}
C_\pm &=&\partial_n U_\pm +\dots, \\
C_{\pm A} &=&\partial_n U_{\pm A} + \dots, 
\end{eqnarray}
and the boundary conditions for either system are
\begin{eqnarray}
\label{BC1}
C_+-\kappa_1C_-&=&0, \\
\label{BC2}
C_{+A}-\kappa_2 C_{-A}&=&0, \\
\label{BC3}
V_+-\kappa_3 V_-&=&F, \\
\label{BC4}
U_{+AB}-\kappa_4 U_{-AB}&=&F_{AB},
\end{eqnarray}
where $F$ and $F_{AB}$ are free boundary data, and
({\ref{BC1}-\ref{BC2}) are implemented as
\begin{eqnarray}
U_++\kappa U_-&=&X, \label{BC1'}\\
U_{+A}+\kappa U_{-A}&=&X_A.\label{BC2'}
\end{eqnarray}

In general the boundary system is coupled to the bulk system, so that
$X$ and $X_A$ in ({\ref{BC1'}-\ref{BC2'}) cannot be considered as
given a priory, and then the constraint-preserving boundary conditions
are not true maximally dissipative
boundary conditions. There are two exceptions, which have been called
``Neumann'' and ``Dirichlet'' boundary conditions in the literature,
where an extended boundary system can be given that decouples from the
bulk system. This happens for the Einstein-Christoffel system
linearised around Minkowski spacetime, \cite{CalabreseCPBC}, the full
Einstein equations in harmonic gauge \cite{SzilagyiWinicour}, and the
Maxwell equations \cite{bssn1}. This boundary system can then be
evolved before the bulk system is evolved,  $X$ and $X_A$ can 
be treated as given a priori, and the constraint-preserving
boundary conditions become true maximally dissipative boundary
conditions. The details, assuming a smooth boundary with tangential
shift, are given in Appendices~\ref{appendix:dirichlet} and
\ref{appendix:neumann}. However, we would like to stress that these
boundary conditions are very restrictive: it does not seem very
physical to find the boundary of the spacetime without knowing what is
inside.

As we have assumed that the shift is everywhere tangential to the
boundary, and this is possible in the case of a non-smooth boundary
(for example a cube) only for zero shift, we also restrict the
analysis in this paper to a smooth boundary. 


\subsection{Mode analysis}


If the boundary system does not decouple, we cannot use our current
energy estimates to prove well-posedness of the initial-boundary value
problem. We can, however, check a necessary condition for
well-posedness, namely that there are no modes that grow exponentially
in time where the growth rates increases unboundedly with spatial
frequency. We conjecture that this condition is also sufficient.

In the frozen coefficients approximation that we have been using
throughout this paper we assume that the linearised perturbation
varies over space and time scales much smaller than those given by the
background solution and the numerical domain. For consistency we must
therefore assume that the domain is a half space and that the boundary
is a plane. We introduce coordinates so that the domains is
$-\infty<x^1\le 0$ and $-\infty<x^A\equiv(x^2,x^3)<\infty$, and the
metric in these coordinates is $\delta_{ij}$. In the frozen
coefficient approximation $\alpha>0$ and $\beta^i$ are also constant in
space and time. (As before we assume $\beta_n=0$ on the boundary, and
therefore everywhere in the frozen coefficient approximation). 

After a Fourier transform in $x^A$ and Laplace transform in $t$ we are
left with a system of linear ODEs with constant coefficients in
$x^1$. In general this can be transformed into a matrix eigenproblem
by an exponential ansatz in $x^1$. (A priori an exponential times a
polynomial in $x^1$ could be required to find the general solution but
this is not the case here.)  The general solution with homogeneous
boundary conditions can therefore be written as a sum of modes of the
form
\begin{equation}
\label{LapFour}
u(x^1,t)=e^{(\alpha s+i\beta^A\omega_A)t
+i\omega_Ax^A+\mu  x^1}\bar u
\end{equation}
where $\bar u$ is a constant vector. With this ansatz we
have $\p_0 u=su$, $\p_n u=\mu u$ and $\p_A u=i\omega_A u$.

If the initial data are nonzero, we must add to the sum over modes a
particular integral that does not concern us here because it is
controlled by the boundary data \cite{GustafssonKreissOliger}.  We are
interested only in modes that are square-integrable over space at any
moment in time. Therefore we assume that $\omega_A$ is real, and that
$\re\mu>0$. $s$ and $\mu $ will in general be complex. If a mode of
this form exists for some $(s,\omega_A,\mu ,\bar u)$, then one exists
also for $(ks,k\omega_A,k\mu,\bar u)$ for any $k>0$. Therefore, if any
growing mode, with $\re s>0$, exists, there are growing modes with
arbitrarily large growth rates and the problem is ill-posed. A
necessary condition for well-posedness of the initial-boundary value
problem is therefore that the homogeneous boundary conditions rule out
the existence of any mode with $\re s>0$ for real $\omega_A$ and $\re
\mu>0$.

For simplicity we concentrate again on NOR with
(\ref{NORexample1}-\ref{NORexample3}). $u=(\g_{ij},K_{ij},f_i)$ is
decomposed with respect to the normal vector $n_i=(1,0,0)$. It is
helpful to introduce the notation $f_m\equiv i\omega^A f_A$ and
$f_p\equiv p^if_i$ where $p^i$ is orthonormal to $\omega_i$ and $n_i$,
and similarly for other tensor components. Substituting the ansatz
(\ref{LapFour}) into the NOR evolution equations, we find after some
linear algebra that, for $s\ne 0$,
\begin{equation}
\label{scondition}
(s^2+\omega^2-\mu ^2)\bar\g_{ij}=0
\end{equation}
and 
\begin{eqnarray}
\label{Kbar}
\bar K_{ij}&=&-{s\over 2} \bar\g_{ij}, \\
\bar f_p&=&0, \\
\bar f_m&=&-{2\over 3}\omega^2(\bar\g_{nn}+\bar\g_{qq}), \\
\label{fnbar}
\bar f_n&=&{2\over 3}\mu (\bar\g_{nn}+\bar\g_{qq}).
\end{eqnarray}
For a non-zero solution to exist, we must have
\begin{equation}
\label{c1}
\mu^2-s^2=\omega^2.
\end{equation}
The coefficients $\bar\g_{ij}$ are then free parameters. They determine
the coefficients $\bar K_{ij}$ and $\bar f_i$ through
(\ref{Kbar}-\ref{fnbar}).

Similarly, for the BSSN system with
(\ref{BSSNexample1}-\ref{BSSNexample2}) we find the equivalent of
condition (\ref{scondition}) for $\bar{\tg}_{ij}$ and $\bar\phi$, and
\begin{eqnarray}
\label{KbarBSSN}
\bar{K} &=& -6s\bar\phi , \\
\bar{\tA}_{ij} &=& -\frac{s}{2} \bar{\tg}_{ij} , \\
\bar{\tG}_p &=& 0 , \\
\bar{\tG}_m &=& -8\omega^2\bar\phi , \\
\bar{\tG}_n &=& 8\mu\bar\phi .
\end{eqnarray}
Note that $s\bar{\tg}_{ij}$ represents the time derivative of
$\tg_{ij}$. Therefore, as we assume that the algebraic constraint
$D=0$ is being imposed continuously, $\bar{\tg}_{ij}$ must be
tracefree. Similarly, from the algebraic condition $T=0$,
$\bar{\tA}_{ij}$ is also tracefree.  With (\ref{scondition}) obeyed,
$\bar{\tg}_{ij}$ and $\bar\phi$ are free coefficients, and determine
$\bar{\tA}_{ij}$, $\bar K$ and $\bar{\tG^i}$.

We now substitute the ansatz (\ref{LapFour}) with these coefficients
into the six constraint equations
(\ref{BC1}-\ref{BC4}). We obtain six algebraic
equations that are linear in the four components $\bar\g_{nn}$,
$\bar\g_{nm}$, $\bar\g_{np}$, $\bar\g_{qq}$ of $\bar\g_{ij}$ and the
two components, $\bar\g_{mm}\equiv-\bar\g_{pp}$, $\bar\g_{mp}$ of the
tracefree transversal object $\bar\g_{AB}$. We can solve these
recursively to find that all $\bar \g_{ij}=0$, as long as
\begin{equation}
\label{c2}
(1-\kappa_i)\mu +(1+\kappa_i)s\ne 0.
\end{equation}
for all four $\kappa_i$. For a mode to exist, this inequality must be
violated for at least one of the $\kappa_i$. Let the value of this
$\kappa_i$ be $\kappa$, which therefore obeys
\begin{equation}
\label{c3}
\kappa(\mu-s)=(\mu+s).
\end{equation}

We now investigate the space of possible solutions
$(\mu,s,\omega,\kappa)$ of the two algebraic equations (\ref{c1}) and
(\ref{c3}) with $\re \mu>0$ and $\re s>0$, with the aim of finding a
condition on $\kappa$ that excludes all such solutions.
We first consider the case $\omega=0$. Then either $\mu=s=0$, or
$\mu=-s$ and $\kappa=0$. Either solution does not correspond to
growing square-integrable modes. We can now assume $\omega>0$, and 
parameterise all solutions by $s$. We find
\begin{equation}
\label{mukappa}
\mu(s)=\sqrt{s^2+\omega^2}, 
\qquad \kappa(s)=\left({\mu-s\over\omega}\right)^{-2}.
\end{equation}
We choose the principal branch of the square root, because it
maps $\re s>0$ to $\re \mu>0$, that is, the growing modes are precisely
the square-integrable modes. This choice also maps $\re s>0$ to
$|\kappa|>1$. Therefore, we exclude all growing square-integrable
modes if we restrict $|\kappa_i|\le 1$ for all four $\kappa_i$, or 
$-1\le\kappa_i\le 1$ for real $\kappa_i$. More details are given in
Appendix~\ref{appendix:kappa}. 


\section{Conclusions}
\label{section:conclusions}


We have constructed families of generalisations of the BSSN and NOR
variants of the ADM evolution equations that are symmetric hyperbolic
in the sense defined for second-order systems in \cite{bssn1}. This
confirms the previous result of \cite{SarbachBSSN} on the BSSN
equations, without recourse to any first-order reduction, and
generalises it by finding the most general energies for the main and
constraint system. This generalisation allows all characteristic modes
to propagate with causal speeds, in particular with speeds $(0,\pm 1)$
only. 

In our analysis of the BSSN equations we also clarify the role for
hyperbolicity of imposing the algebraic constraints $\det\g_{ij}=1$
and ${\rm tr}\tA_{ij}=0$ during the evolution. We find that the
equations can be made symmetric hyperbolic if these constraints are
imposed continuously, and strongly hyperbolic without imposing the
constraints, but adding an extra term to the evolution equations.

There is numerical evidence that densitising the lapse and imposing
the trace constraint improves stability in moving single black hole
simulations, even without imposing maximally dissipative
constraint-preserving or boundary conditions
\cite{LagunaShoemakerBSSN}. This is not surprising, as these changes
make the evolution equations strongly hyperbolic, and imposing the
determinant constraint as well would make them symmetric
hyperbolic. 

Our results go some way towards explaining why the BSSN system has
been relatively successful in simulating black hole or neutron star
binaries. It is possible that the Bona-Mass\'o formulation \cite{BM},
a strongly hyperbolic first-order version of the Einstein equations
that introduced variables similar to the $\tG^i$ or $f_i$, has not
been as successful because it is first order, which we expect makes it
more susceptible to constraint-violating instabilities of the
convergent type.

The NOR system is basically the BSSN system without the
conformal-traceless decomposition, and the similarity of our results
for the two systems suggests that the NOR system shares all the
advantages of the BSSN system, without the overhead of the extra
variables $K$ and $\phi$ and extra constraints $T=0$ and $D=0$.

With symmetric hyperbolicity, we can make the initial-boundary value
problem formally well-posed by imposing maximally dissipative boundary
conditions. However, these boundary conditions are in general not
compatible with the constraints, and so large constraint violations
(of the convergent type) propagate in from the boundaries. This can be
avoided by replacing some of the maximally dissipative boundary
conditions on the main system by maximally dissipative boundary
conditions on the constraint system \cite{CalabreseCPBC}, and we have
given details of how to do this for NOR and BSSN, for the case of a
smooth boundary with tangential shift. Note that even when we have
fixed the principal part of the field equations both the main and
constraint energies still depend on a number of free parameters, which
appear explicitly in the boundary conditions.

Except for two rather unphysical special cases, we have not proved
that the initial-boundary value problem with constraint-preserving
boundary conditions is well-posed. We have shown, however, that these
boundary conditions rule out perturbation modes with unbounded growth,
which is a key necessary condition for well-posedness
\cite{GustafssonKreissOliger}. We plan to investigate a proof of
well-posedness, and in parallel to examine stability in numerical
experiments.

Our discussion of both maximally dissipative and constraint-preserving
boundary conditions assumes that the normal component $\beta_n$ of the
shift vanishes at the boundary, as then the $\lambda=0$ characteristic
modes propagate along the boundary. This restriction allows
for a shift that is everywhere tangential to a smooth boundary, and
this could be used for example to employ corotating coordinates in the
simulation of a binary system. The case of a general shift will be
investigated in future work.

All equations in this paper were derived using {\em xTensor}, an
open-source {\em Mathematica} package for abstract tensor calculus,
developed by JMM. It is available under the GNU Public Licence from
{\tt http://metric.imaff.csic.es/Martin-Garcia/xAct/}.


\acknowledgments


The authors would like thank G Nagy, O Ortiz and O Reula for
communicating a draft paper and for discussions. CG would like to
thank the Kavli Institute for Theoretical Physics for hospitality
while this work was begun, and T Baumgarte, G Calabrese, H Friedrich,
L Kidder, L Lindblom, O Sarbach, M Scheel, D Shoemaker and J Vickers
for helpful discussions.
JMM was supported by the Comunidad Aut\'onoma de Madrid and Fondo
Social Europeo and also in part by the Spanish MCYT under the
research project BFM2002-04031-C02-02.


\begin{appendix}


\section{The KST formulation}
\label{appendix:kst}


The Kidder-Scheel-Teukolsky (KST) formulation \cite{KST} is based on a
reduction to first order of the ADM evolution equations with a
densitised lapse (DADM) with the auxiliary variables $d_{kij}\equiv
\p_k\g_{ij}$. This gives rise to the auxiliary constraints
$C_{ijkl}\equiv \p_{[i}d_{j]kl}=0$. The principal part of the evolution
equations is
\begin{eqnarray}
\p_0 \g_{ij}&\simeq&-2K_{ij}, \label{KSTgdot}\\
\p_0 K_{ij}&\simeq& \hbox{DADM} + \gamma\, \g_{ij} H + \zeta\, \g^{kl}
C_{k(ij)l}, \\
\p_0 d_{kij} &\simeq& \hbox{DADM} + \eta\, \g_{k(i}M_{j)}+\chi \,
\g_{ij} M_k. \label{KSTddot}
\end{eqnarray}
This system can be made strongly or symmetric hyperbolic for certain
ranges of the parameters $\sigma$, $\gamma$, $\zeta$, $\eta$ and
$\chi$. In particular, the Einstein-Christoffel (EC) system is
the case $\gamma=0, \zeta=-1, \eta=4, \chi=0$, densitizing the lapse
with $\sigma=1$ in our notation.
In this Appendix we want to point out that only
$\sigma$ and $\gamma$ have counterparts in a second-order
system. $\zeta$ has a similar function to our parameter $a$, and
$2\eta+\chi$ has a similar function to our parameter $b$, but these
parameters vanish if we replace $d_{kij}$ by $\p_k\g_{ij}$: $C_{ijkl}$
then vanishes identically, and $d_{ijk}$ is no longer evolved
explicitly by (\ref{KSTddot}), but only implicitly by
(\ref{KSTgdot}). Comparing the KST system to NOR with the benefit of
hindsight, one could say that the only indispensable effect of
$d_{kij}$ is to introduce the divergence ${d^k}_{kj}$ as an auxiliary
variable.


\section{Dirichlet boundary system}
\label{appendix:dirichlet}


With 
\begin{equation}
-\kappa_1=\kappa_2=\kappa_3=\kappa_4=1
\end{equation}
we have (in our example NOR system)
\begin{eqnarray}
F&=&2K_{nn}, \\
F_{AB}&=&-2 K_{AB}, \\
X&=&4K_{qq}, \\
X_A&=&2(f_A-\pn \g_{An})+(\rho-2)\p_A(\g_{nn}+\g_{qq}).
\end{eqnarray}
The boundary system 
\begin{eqnarray}
\p_0 X&=&\p^AX_A-2\p^A(\p_A \g_{qq}), \\
\p_0 X_A&=&\frac{1}{2}\p_A X+2\p_A F+2\p^B F_{AB}, \\
\p_0 (\p_A \g_{qq})&=&-{1\over 2}\p_A X
\end{eqnarray}
decouples from the bulk system. Note that all variables of the
boundary system have parity $+1$ under the reflection $n_i\to -n_i$
through the boundary. The boundary system is strongly hyperbolic with
characteristic variables
\begin{eqnarray}
X_\pm&=&4\p_m\g_{qq}-2X_m\mp\sqrt{6}X, \\
X_0&=&X_m+\p_m\g_{qq}, 
\end{eqnarray}
which have speeds $\lambda=(\pm\sqrt{3/2},0)$, as well as $X_p$ and
$\p_p\g_{qq}$ with zero speed. (It is also symmetric hyperbolic,
but this is not required for well-posedness if the boundary is smooth
without boundary, and therefore we do not give details here.)


\section{Neumann boundary system}
\label{appendix:neumann}


With 
\begin{equation}
-\kappa_1=\kappa_2=\kappa_3=\kappa_4=-1
\end{equation}
we have
\begin{eqnarray}
F&=&2f_n-\pn\g_{nn}+(\rho-2)\pn(\g_{nn}+\g_{qq}), \\
U_0'&=&f_n+\frac{\rho-2}{2}\pn(\g_{nn}+\g_{qq}), \\
F_{AB}&=&\pn \g_{AB}, \\
X&=&-2\pn \g_{qq}, \\
X_A&=&4K_{An}.
\end{eqnarray}
The autonomous boundary system is 
\begin{eqnarray}
\p_0 X&=& \p^A X_A, \\
\p_0 X_A&=&\frac{1}{2}\p_A X +2\p^B F_{AB} + 2\p_A F
\nonumber \\
&&-2\p_A U_0' -2\p^B\p_B \g_{An} , \\
\p_0 (\p_A \g_{Bn})&=&-{1\over 2}\p_A X_B, \\
\p_0 U_0' &=& 0 . 
\end{eqnarray}
Note that all variables of the boundary system have parity $-1$.
The boundary system is strongly hyperbolic with characteristic variables
\begin{eqnarray}
X_\pm&=&4\p_m\g_{mn}-X\mp\sqrt{6}X_m, \\
Y_\pm&=&\p_m\g_{pn}\mp{1\over 2}X_p, \\
X_0&=&\p_m\g_{mn}+{1\over 2}X, 
\end{eqnarray}
with speeds $\lambda=(\pm\sqrt{3/2},\pm 1, 0)$, as well as the zero
speed variables $U_0'$, $\p_p\g_{mn}$ and $\p_p\g_{pn}$.


\section{Details of the mode analysis}
\label{appendix:kappa}


\begin{figure*}
\includegraphics[width=15cm]{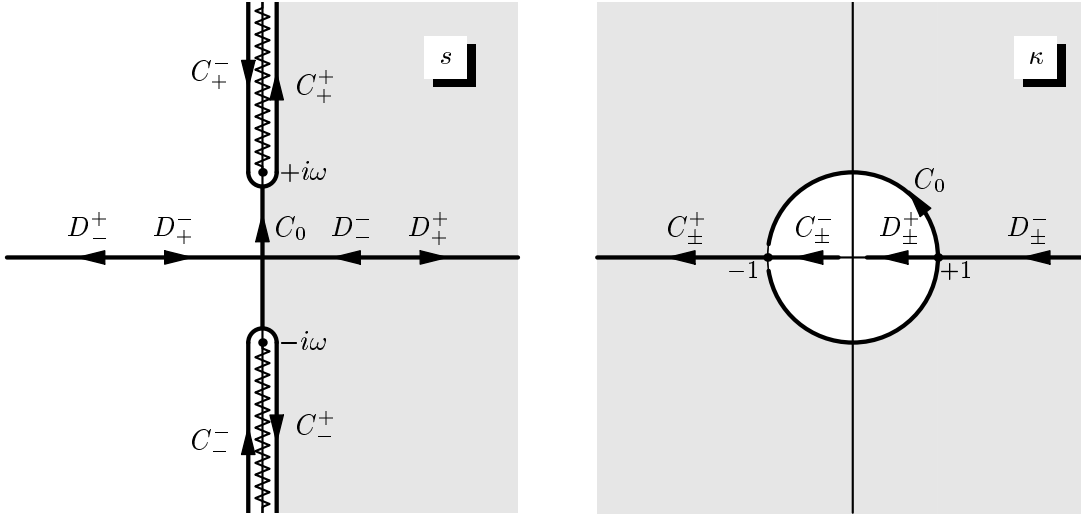}
\caption{\label{contours} Contours $C$ and $D$ in the complex $s$
plane and their images in the $\kappa$ plane under (\ref{mukappa}).
Arrows point towards increasing parameter $\varphi$.  The shaded area
in the $s$ plane is mapped to the shaded area in the $\kappa$ plane.
}
\end{figure*}

An alternative approach to finding the range of real values of
$\kappa$ that exclude growing modes would be to solve (\ref{c1}) and
(\ref{c3}) explicitly for the values of $s$ and $\mu$ that are allowed
for a given $\kappa$, and then check if they correspond to a growing,
square-integrable mode. For $\kappa=0$ we have $\omega=0$ and
$\mu=-s$. Therefore $\re\mu>0$ implies $\re s<0$, and vice versa,
and so there are no growing square-integrable modes. For each
$\kappa\ne 0$ we find two values of $s$ and $\mu$. For $\kappa>0$, we
can parameterise them as
\begin{eqnarray}
D_\pm: \quad s&=&\pm\omega\sinh\varphi, \qquad
-\infty<\varphi<\infty, \nonumber \\ 
\Rightarrow \mu&=&\pm\omega\cosh\varphi, \qquad
\kappa=e^{-2\varphi}.
\end{eqnarray}
The solution $D_+^+$ with the upper sign and $\varphi>0$ has $\re s>$
and $\re \mu>0$, and corresponds to growing square-integrable
modes. To exclude these, we must demand $\kappa\le1$. However, for
$\kappa<0$ there is a potential fallacy in this purely algebraic
approach.

Consider the complex $s$ plane. For definiteness we place branch cuts
from $i\omega$ to $i\infty$ and from $-i\omega$ to $-i\infty$.
Consider the two contours $C_+$ and $C_-$, parameterised by $\varphi$,
which wrap around the upper and lower branch cut respectively:
\begin{eqnarray}
C_\pm: \quad s&=&\pm i\omega\cosh\varphi, \qquad
-\infty<\varphi<\infty, \nonumber \\
\Rightarrow \mu&=&\pm i\omega\sinh\varphi, \qquad
\kappa=-e^{2\varphi}.
\end{eqnarray}
These give us the two possible values of $\mu$ and $s$ for each real
$\kappa<0$. From a point of view where we only consider real values of
$\kappa$, none of these modes are growing or square-integrable, and so
we do not seem to obtain any restriction on negative
$\kappa$. However, we should consider modes with purely imaginary
$\mu$ and $s$ as limiting cases of growing square-integrable modes as
$\re s\to 0_+$ and $\re\mu\to 0_+$. Clearly this is the case for the
modes lying on the contours $C_\pm^+$ (i.e. $\varphi>0$) which are to
the right of the branch cuts and therefore contiguous with $\re s>0$,
but not for the contours $C_\pm^-$, which are to the left of the
branch cut. For the modes on $C_\pm^+$, $|\kappa|>1$, and so they are
suppressed by any boundary condition with $|\kappa|\le 1$. This is the
range of $\kappa$ that we expect from the energy method. Another way
of seeing that the modes on $C_\pm^-$ should not be considered is to
note that our mode analysis is really derived from a Laplace transform
in $t$ \cite{GustafssonKreissOliger}, so that we need a contour for
the inverse Laplace transform that is to the right of all branch cuts
(and singularities) in the $s$ plane.

The boundary of $\re s>0$ is given by the union of the three contours
$C_-^+$, $C_0$ and $C_+^+$ where 
\begin{eqnarray}
C_0: \quad s&=&i\omega\sin\varphi, \qquad
-{\pi\over 2}<\varphi<{\pi\over 2} ,\nonumber \\
\Rightarrow\mu&=&\omega\cos\varphi, \qquad 
\kappa=e^{2i\varphi}
.
\end{eqnarray}
This contour can also be used as a contour for the inverse Laplace
transform. All contours in the $s$-plane discussed here and their
images in the $\kappa$ plane are shown in Fig.~\ref{contours}. The
shaded areas are $\re s>0$ and its image $|\kappa|>1$. $\re s$ is
bounded by by the union of the three contours $C_-^+$, $C_0$ and
$C_+^+$ and $\re \mu$ by their images. This proves the claim made
above that $\re s>0$ is mapped to $|\kappa|>1$.


\end{appendix}



\end{document}